\documentstyle[12pt]{article}

\begin{document}

\title{Golden Section and the Art of Painting}
\author{Agata Olariu\\
{\em National Institute for Physics and Nuclear Engineering}\\
{\em P.O.Box MG-6, 76900 Bucharest Magurele, Romania}}
\maketitle

\section{Introduction}

In mathematics, the {\it Golden Section} is a geometric proportion created by a
point C on a segment of line AB when AC/AB=CB/AC, as shown in Fig. 1.\\
This ratio has the value $\Phi$=0.618...\\
Since the times of antiquity many philosophers, artists and mathematicians
have been preoccupied by the Golden Section, which the writers of the
Renaissance have called the "Divine Proportion". The mathematician 
Lucas Pacioli has characterized the Golden Section as aesthetically satisfying
and wrote on this theme the treaty "Divina Proportione". 
It is largely accepted that a rectangle having the sides in this ratio has
special aesthetic qualities $^{1}$. Moreover the Golden Section has been used
as an ideal proportion on which the pattern of lines and shapes in the
composition of a painting should be based.
Taking this idea as a point of departure in this paper was done a statistical
study on a series of paintings, belonging to various authors and from 
different periods to see how
the Golden Section is applied in painting. The ratio
between the sides of the paintings chosen by these painters
is regarded as the most appropriate and beautiful proportion. 
Let's remark that 1/$Phi$=1.618 is also related to the
golden section by the relation $1/\Phi=\Phi +1$.

\section{Statistical study of paintings}

It was done a statistical study  on 565 works of art of 
different great painters: Bellini, Caravaggio, Cesanne$^2$, Goya, van Gogh, 
Delacroix, Pallady (Romanian painter), Rembrandt, Toulouse-Lautrec.
It was calculated the ratio of the 2 dimensions of painting: the longer part
 to the shorter part of the painted rectangle.
In Table 1 are given the average values and the errors of the average for the 
ratio of the sides for various paintings of the painters studied. The paintings
considered in this statistics have been selected from the specified references,
where the sides of the paintings have been indicated.

\small
\begin{table}[h]
\newlabel{}
\caption{{\bf Tabelul 1}. Average of the ratio of the sides of paintings,
together with the error of the average for a number of paintings belonging to
various painters}\\

\begin{tabular}{lcc}
\hline
Painter   &          Number of paintings & Average L/l, error\\
          &             considered       &                  \\
\hline
\hline\\
Bellini (Venetian )    & 53             & 1.46 $\pm$0.10\\
Caravaggio             & 37             & 1.32 $\pm$0.15\\              
Cezanne                & 100            & 1.26 $\pm$0.27              \\
Delacroix              & 42             & 1.32 $\pm$0.17 \\
Van Gogh               & 69             & 1.32 $\pm$0.19               \\
Goya                   & 34              & 1.04 $\pm$0.04           \\
Pallady                & 127            & 1.30 $\pm$ 0.16\\
Rembrandt              & 39             & 1.33 $\pm$0.14 \\
Toulouse-Lautrec       & 64             & 1.36 $\pm$0.12\\
\hline
\end{tabular}
\end{table}

Assuming that all the painters under discussion enter in a statistics with
equal weights, in Fig. 2a is shown the total distribution, for the number of
paintings N= 565. The average value obtained for the ratio of the sides is
\begin{center}
 1.34 $\pm$ 0.12. 
\end{center}
This value, determined experimentally, is the result of the intuitive choice of
great creators of art and is significsantly different from the value of the
Golden Section $1/\Phi$=1.618, which is a theoretical ratio, obtained from an
abstract, mathematical theory, which supposedly ought to impress on a painting
a supreme harmony.\\

In Fig. 2b is  illustrated the ratio L/l=1.34, lying on the maximum
of the distribution, by the painting of Toulouse-Lautrec, "La Goulue entering
at Moulin Rouge", dated 1891-1892, Museum of Modern Art from New York, 
having the dimensions 79.4 x 59 cm, and alongside this painting it is  drawn 
a rectangle whose sides are in the Golden Section.\\

\vspace*{1cm}

{\bf \large References}\\
\noindent
1. Peter B. Norton, Josph J. Esposito, The New Encyclopaedia Britannica, \\
\hspace*{2cm}15$^{th}$ Edition, 1995 
2. Nicolas Pioch, WebMuseum Data Base \\
\end{document}